\documentclass[12pt,preprint]{aastex}

\newcommand{\kms}{$\rm km\,s^{-1}$}

\shorttitle{HII regions in NGC 3389}
\shortauthors{Abdel-Hamid et al.}

\begin{document}

\title{HII regions in the spiral galaxy NGC 3389} 
\author{Hamed Abdel-Hamid$^{1,2}$ and Sang-Gak Lee$^{1}$}
\affil{1- Astronomy Program, SEES, Seoul National University, Seoul 151-742, Korea}
\affil{2- National Research Institute of Astronomy and Geophysics, Cairo, Egypt}
\email{hamed@astro.snu.ac.kr, sanggak@astrosp.snu.ac.kr}
\and
\author{Peter Notni}
\affil{  Astrophysikalisches Institut Potsdam, D-14482 Potsdam, Germany}
\email{pnotni@aip.de}


\begin{abstract}
CCD observations in V, I and H$_\alpha$ for NGC 3389 are used to present photometry of 61 HII regions. Their $\alpha$ and $\delta$ positions, diameters and their absolute luminosities have been determined. The luminosity and size distribution functions of the HII regions in NGC 3389 are discussed.
\end{abstract}


\section{Introduction}
NGC 3389 is a spiral galaxy, classified by de Vaucouleurs et al. (1991) as SA galaxy and by Sandage and Tammann (1981) as Sc galaxy.
NGC 3389 is one of the Leo triplet with the galaxies NGC 3384 and NGC 3379. 
The lack of detailed photometric studies for NGC 3389 and its normal undistorted morphology, although it lies in a group of galaxies, motivate us to do the photometry of NGC 3389 and study its stellar population properties.
The basic parameters of NGC 3389 are given in the Table \ref{par}. 

HII regions in galaxies are powered by the newly formed stars, therefore they give
the clues for the recent star formation in galaxies. The total H$_\alpha$
luminosity can be used to quantify their current star formation rate. 
Therefore we concern firstly  with the investigation of the HII region 
properties of NGC 3389; their luminosity and size distributions.

This is the first CCD work on the HII regions of NGC 3389.   
Hodge and Kennicutt (1983) presented the off-set positions for 50 identified HII regions in NGC 3389 in their photographic surveys of the HII regions in galaxies, while Kennicutt and Kent (1983) gave the total integrated H$_\alpha$ flux and luminosity for the galaxy. 
\section{Observations and data reduction}

The V, I and H$_\alpha$ CCD observations of NGC 3389  were obtained using
1.23 m telescope of Calar Alto observatory on 5-6 February 1995
with a  Tek CCD with 1024 $\times$ 1024 pixels with a pixel scale
of 0.502$^{\arcsec}$ pixel$^{-1}$. The sky was photometric with seeing varying from 0.7$^{\arcsec}$ to 1.2$^{\arcsec}$ during the night.
Four H$_\alpha$ frames were taken with H$_\alpha$ interference filter of central bandpass at 6580 $\AA$ and a FWHM of 100 $\AA$.
Four frames in V  filter and three frames in I filter were observed. Table \ref{obser} gives the log of the observations. 
All data reduction were  done using the image processing program MIDAS. 
The standard image processing routine has been done, then all frames were carefully aligned
 using the position of the common foreground stars in the frames.
The sky background was measured using MIDAS task FIT/FLATSKY with cursor area offset from the galaxy in the frame margin.
After sky subtraction, the cosmic ray effects are removed by median filter.
 A weighted mean image for each filter was computed.
In order to get an accurate continuum level we have choosen two continuum bands on either side of H$_\alpha$; V and I are used. 
The fluxes of the selected stars in the three bands (I, V and H$_\alpha$) were measured, then correlations among them get the H$_\alpha$ continuum level.
Then the H$_\alpha$ image was corrected for the continuum contribution.
For flux calibration, we compare the integrated measurements within 
 an aperture whose diameter equal to three arcminutes, which contains the entire visible disk of NGC 3389 to a limiting surface brightness of 25 mag/arcsec$^{2}$ with the total absolute luminosity given by Bell and Kennicutt (2001). This total integration  normalized by the total absolute luminosity quoted by Bell and Kennicutt (2001),  for the same diameter size (3$^{\arcmin}$), to get the observed measurements in units of erg s$^{-1}$.
 The absolute flux has been measured with the same H$_\alpha$ filter specification ($\lambda$ 6580/100 $\AA$) like that of ours.
Bell and Kennicutt (2001) used to convert the fluxes to luminosity the adopted distance of 24 Mpc refering to Shanks (1997).
\placetable{par}
\placefigure{display}
\section{HII regions identification}
\begin{figure}
\plotone{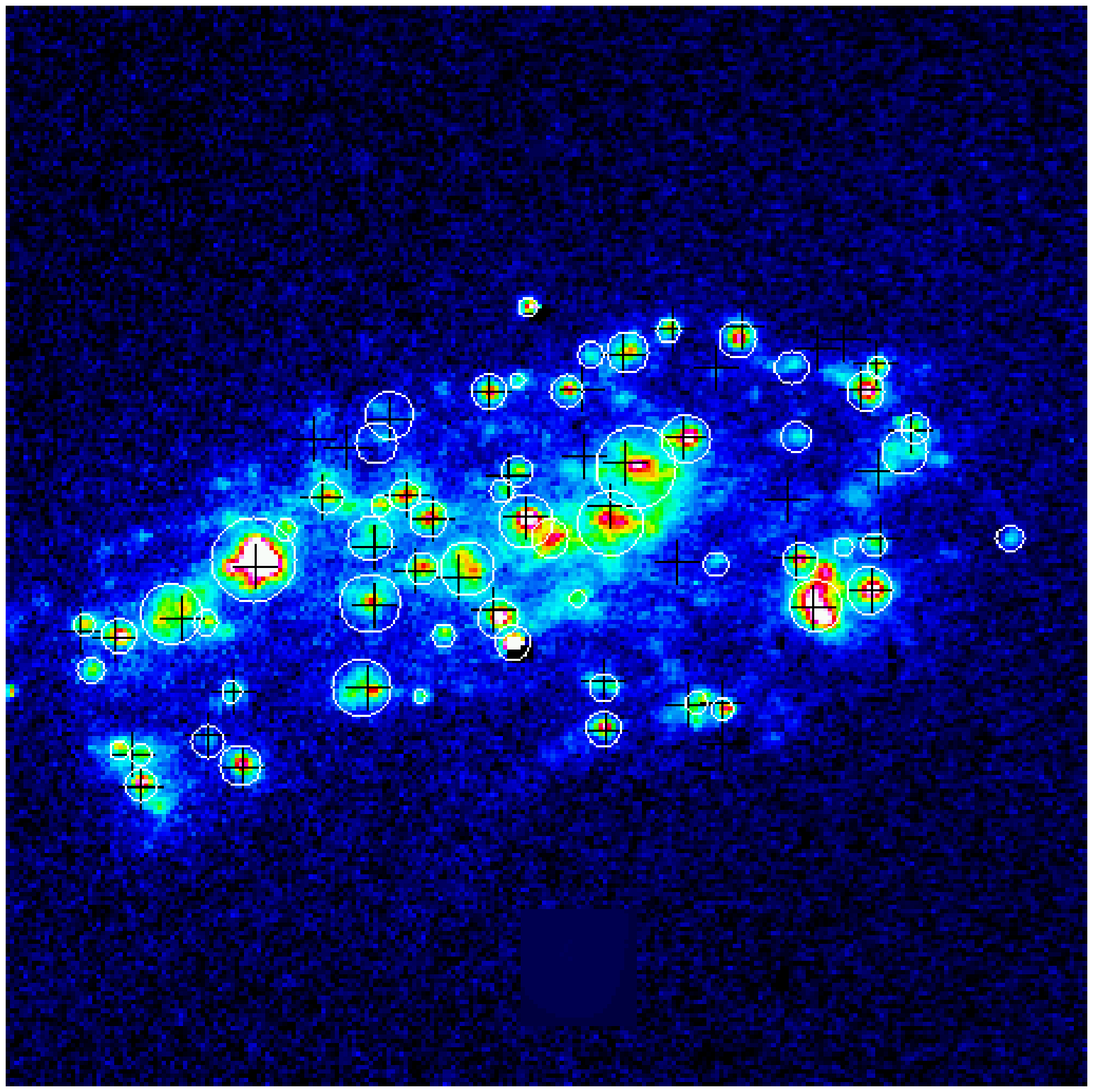}
\caption{Continuum subtracted H$_\alpha$ image for NGC 3389 with HII regions of this study (ellipses) and those of Hodge\& Kennicutt (1983) (crosses). \label{display}}
\end{figure}

The Figure \ref{display} shows the continuum-subtarcted H$\alpha$ image of NGC 3389.
 A visiual identification of the HII regions was firstly done, by displaying
the  continuum-subtarcted H$\alpha$ image with different scaling factors and measure the boundaries at the discernable level.
 To measure the contour dimensions of each  HII region ($dx$ and $dy$) using a graphically interface  precedure within MIDAS was used.      
However if the HII regions are crowded or lie in a diffuse bright background
the largest closed contour is used to measure the HII region boundary. From both methods we fixed dimensions of the HII regions. In some cases it was needed to justifity the aperture radius by hand.
The aperture radius used for the flux integration was calculated as follows:\\
\begin{equation}
r = (\sqrt{dx^{2} + dy^{2}})/2\\
\end{equation}
The aperture radius used
to estimate the backgroud flux in the proximity of each HII region is varied according to how much  HII regions are crowded. 

\begin{figure}
\plotone{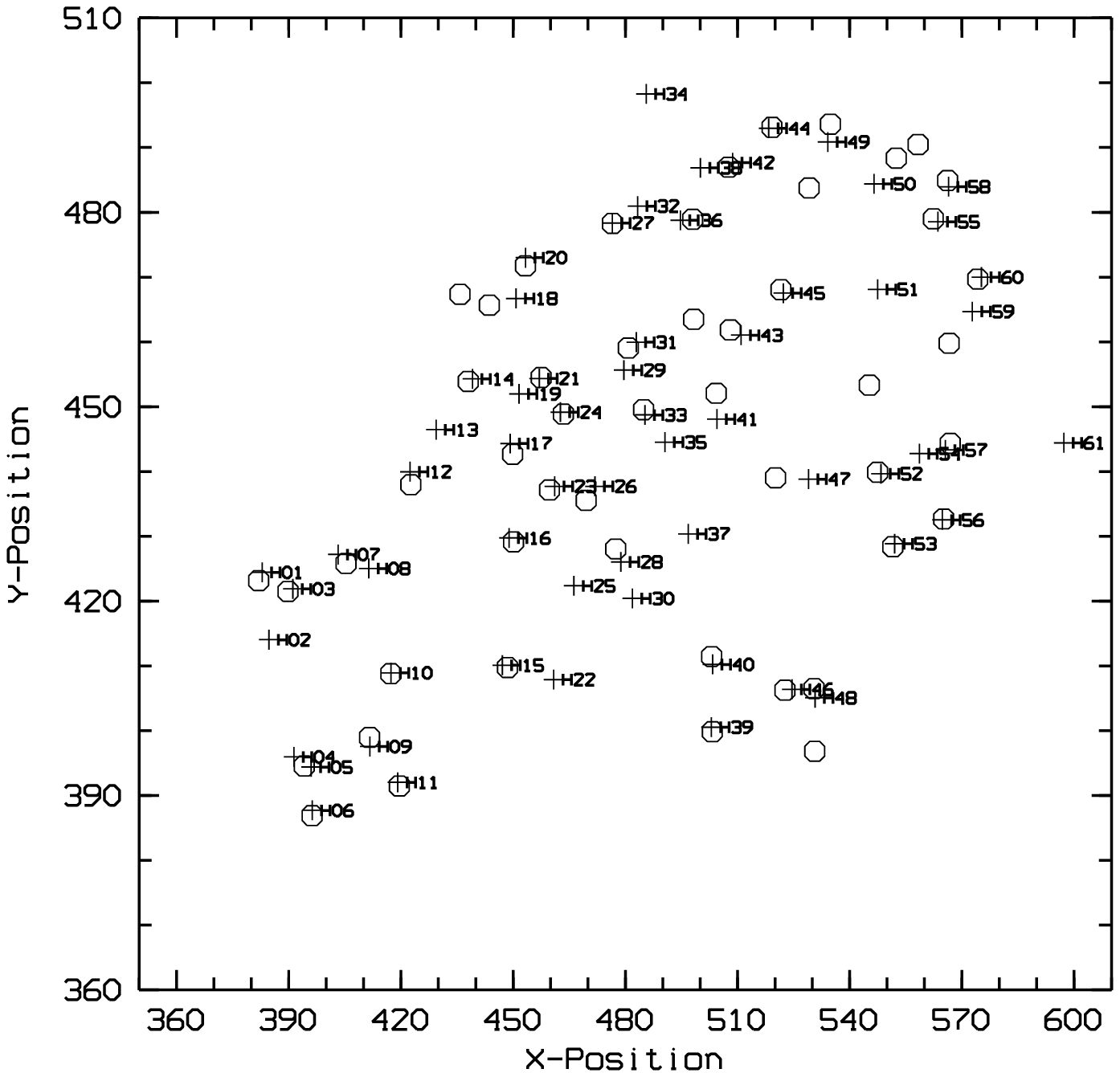}
\caption{Identification of 61  HII regions of this study (open circles) in comparison with those of Hodge \& Kennicutt (1983) (crosses). \label{hod_ham}}
\end{figure}

Figures \ref{display} and \ref{hod_ham} illustrate the positions of 61 HII regions in NGC 3389 in comparison
with those of Hodge and Kennicutt (1983); Hodge provided  the off-set positions of HII regions from the center of NGC 3389 and a map of NGC 3389 by private communication. 
We find a good agreement between our position measurments and their position data for 40 objects. The rest ten HII regions of Hodge and Kennicutt (1983) have no counterparts in our data. These ten HII regions may lie below our identification criterion,
 their off-set positions and identification names are given in Table \ref{Hodge}. Twenty one HII regions are newly identified.

The integrated flux and the background flux for each HII region are measured, then the flux corrected for the background to give the net flux. The main source of error in the net fluxes  
 are due to either the estimation of the background or the inaccurate estimate of the boundaries, or both. To estimate the uncertainties in the flux measurements multiple trials have been done to measure the flux of some individuals HII regions.  We found statistically that the uncertainities in the measured fluxes vary from less than 10$\%$ for the isolated regions to 20$\%$ for the crowded regions.

The measured H$_\alpha$ fluxes are corrected for Galactic extinction, A(0.65${\mu m}$) = 0.073 mag (Schlegel et al. 1998).  
To remove the contamination by the [NII] emission lines in the measured H$_\alpha$ fluxes, the total fluxes were multiplied by a factor of  0.75 according to Kennicutt (1983); the factor gave the average ratio of H$_\alpha$/(H$_\alpha$+[NII]) in the spirals. They have used the same filter bandpass at 6580 $\AA$ and FWHM of 100 $\AA$ in their measurments for galaxies with radial velocities under 3000 \kms as NGC 3389.
Sixty one HII regions were identified, and the equatorial coordinates of their positions, the diameters in parsec, the logarithmical lumnosities and the corresponding Hodge \& Kennicutt (1983) identifications (HK-ID) are listed in Table \ref{HII}.
\section{The Luminosity Function}
\begin{figure}
\plotone{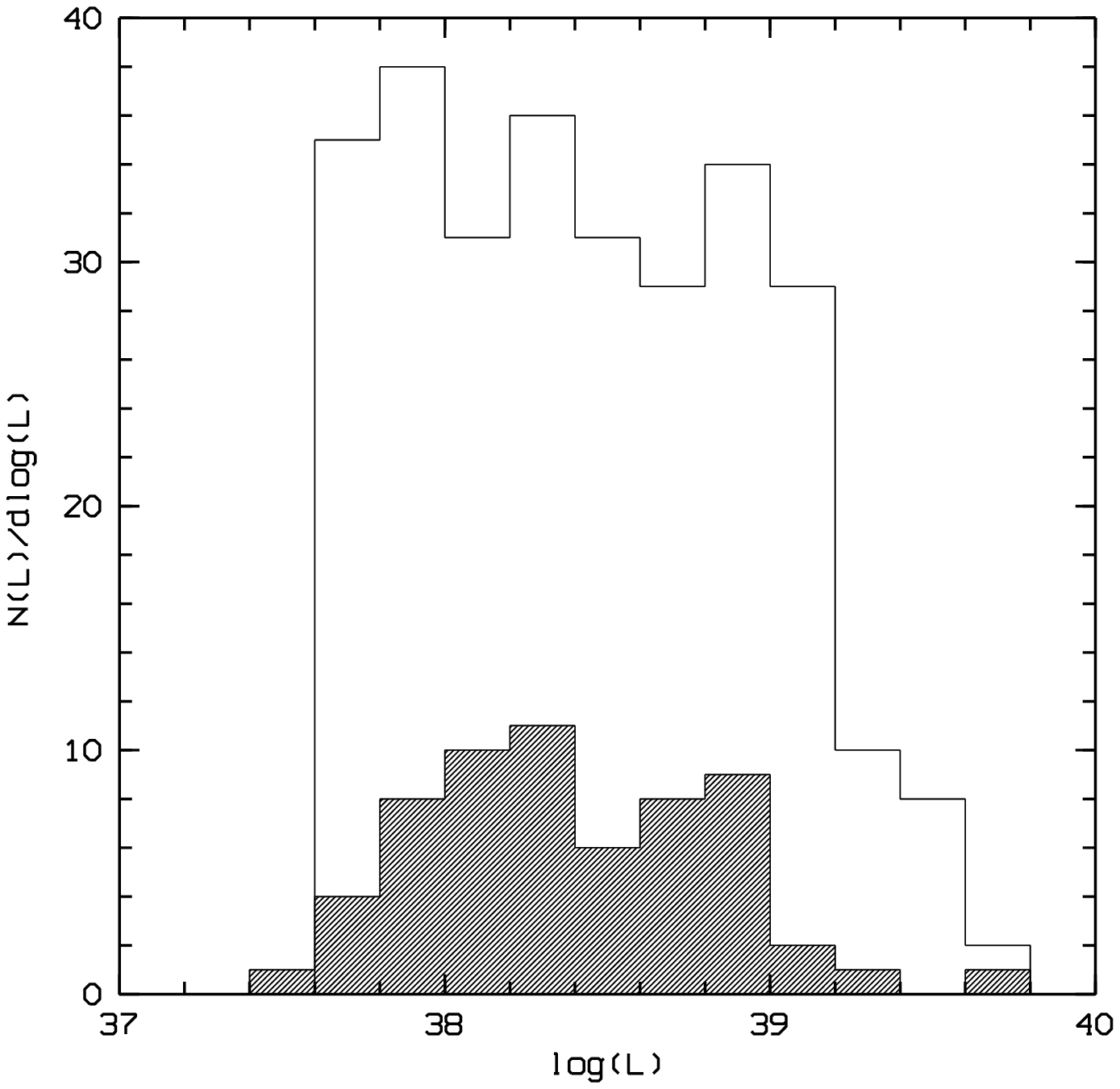}
\caption{The LF of HII regions for NGC 3389 (hashed histogram) against that for NGC 6384 (solid line). \label{lum}}
\end{figure}

The H$_\alpha$ luminosity of the HII regions in NGC 3389 spans a range L(H$_\alpha$) $\approx$ 3$\times10^{37}$-6$\times10^{39}$ erg s$^{-1}$. This L(H$_\alpha$) luminosity range is more frequent in Sa-Sb galaxies (see for example in Thilker et al. 2002). The HII regions with L(H$_\alpha$) luminosity higher than 10$^{39}$ erg s$^{-1}$ are found in late-type (Sc-Irr) normal  galaxies or in interacting galaxies (Kennicutt et al. 1989).
The luminosites of the 61 HII regions are binned in logarithmic interval of 0.2 dex and ploted in Figure \ref{lum}.

In the Figure \ref{lum}, our LF data is compared with that of Feinstein (1997) on the Sb spiral galaxy NGC 6384. The HII regions in both galaxies exhibit nearly  the same L(H$_\alpha$) range, but the number of the HII regions of NGC 6384 is about four times larger than that of NGC 3389. The reason for that is the absolute blue luminosity of NGC 6384 (M$_B$=-21.31) is approximately four times that of NGC 3389. A mimic relation is founded on NGC 6822 by Hodge et al. (1989).

\begin{figure}
\plotone{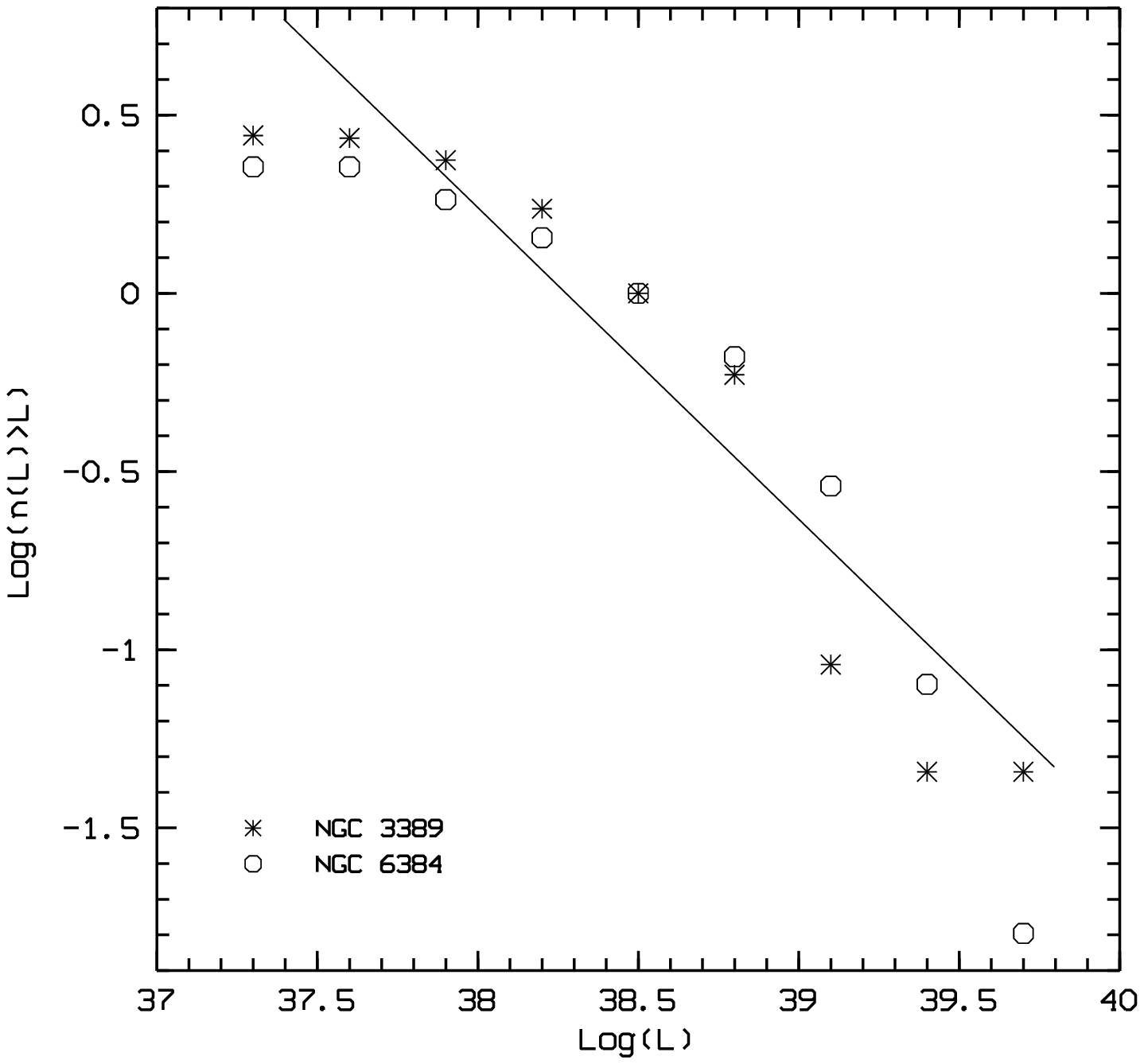}
\caption{The cumulative LF of HII regions on NGC 3389 (stars) in comparsion with that of NGC 6384, (Feinstein 1997) (circles). Solid line is the fitting line for all data points.\label{accu}}
\end{figure}

In order to fit the luminosity distribution of the HII regions independantly of the choosing the LF bin size, the cumulative luminosity distribution have been done following Feinstein (1997). 
The cumulative LF is presented in Figure \ref{accu} in comparison with that of NGC 6384 (Feinstein 1997), both LFs for the two galaxies normalized by their number of the  HII regions at log(L) of 38.6.
 The LFs are in agreement at the luminosities log(L) $\leq$ 39.0, while they are incongruent at higher luminosities log(L) $>$ 39.0. 
This is due to NGC 6384 has 49 HII regions with luminosities higher than 39.0 (log) while NGC 3389 contains only 4 HII regions.
The two galaxies have different absolute blue luminosities as well as different spiral types.  This result may confirm that NGC 3389 is neither Sc spiral nor an interacting galaxy.

The straight line in the Figure \ref{accu} gives the best fit for all data points of NGC 3389 LF, which correspondes to a power law in the form:
\begin{equation}
dN = AL^{\alpha} dL
\end{equation}
With a power index $\alpha$=-1.95$\pm$0.24, which
 is in agreement with $\alpha$ =  -2.0$\pm$0.2
for Sa-Sb galaxies quoted by Banfi et al. (1993) for a sample of 22 spirals and with 
 $\alpha$ =  -2.0$\pm$0.5 for a sample of 30 spirals (Kennicutt et al. 1989).

\section{Diameter distribution}
\begin{figure}
\plotone{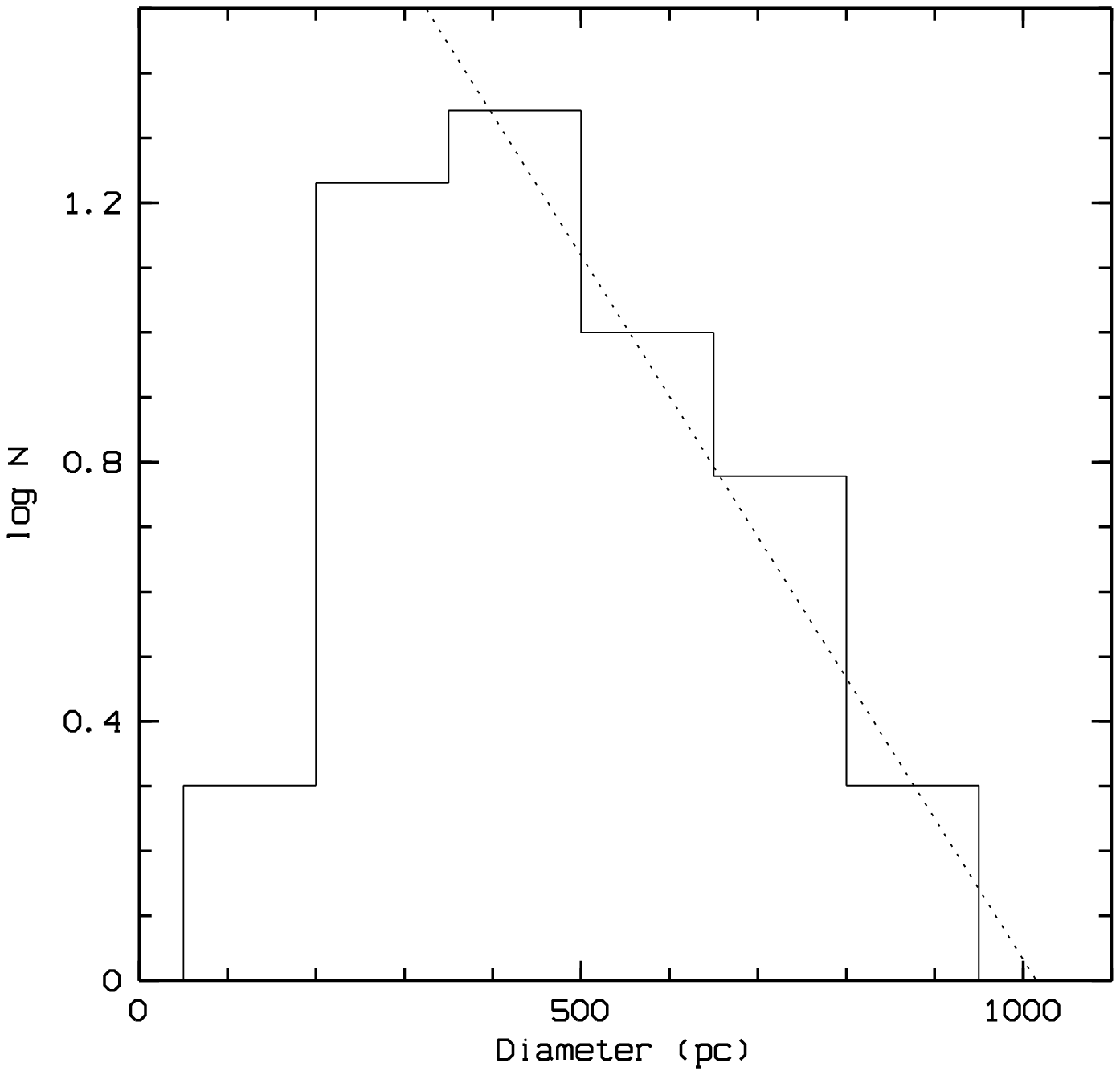}
\caption{Diameter distribution of the HII regions with fitting line (dotted). \label{sizeD}}
\end{figure}

Figure \ref{sizeD} shows the diameter distribution for the HII regions of NGC 3389.
The number of HII regions with diameter greater than a given diameter value plotted as a function of diameter. The data fit an exponential law, as proposed by the 
previous works from Van den Berg (1981), Hodge (1983), Hodge (1987) and many others.
The exponential law in the form N = N$_{\circ}e^{-D/D_{\circ}}$.
We obtain $D_{\circ}$ = 200 $\pm$ 7 pc, which is in agreement with that given by Hodge (1987); 180 pc for NGC 3389, if the smallest diameters were ignored in fitting. 
\section{Summary}
We have identified 61 HII regions in the spiral galaxy NGC 3389. The positions of fourty of them are identical with those in Hodge \& Kennicutt (1983) and  21 are newly identified.
The diameters and the fluxes of 61 HII regions are measured for NGC 3389 for the first time. The H$_\alpha$ luminosities  of HII regions span a range from 3$\times$ 10$^{37}$ to 6 $\times$ 10$^{39}$ erg s$^{-1}$ as those in most Sa-Sb galaxies.  
The luminosity function fits well a power law with a power index is in agreement
with that for early spiral galaxies.
In a forthcoming paper, we will discuss the ionization sources of the HII regions and star formation activity in NGC 3389. 

\acknowledgments
This work is financially supported by SEES-BK21 grant, Korean Government.
Abdel-Hamid is very grateful to BK21 and School of Earth and Environmental Science (Astronomy Program) at Seoul
National University for their support and good hospitality.
We are grateful to P. Hodge for providing us his HII regions data for NGC 3389.

\clearpage

 \clearpage
\begin{table}
\caption{Basic Parameters for NGC 3389 \label{par}}
\vspace*{0.4cm}
\begin{tabular}{lll}
\tableline\tableline
R.A. (2000.0)& 10 ${^h}$ 48 $^{m}$ 28 $^{s}$&\\
Dec  (2000.0)& 12 $^{d}$ 31 $^{\arcmin}$ 59 $^{\arcsec}$&\\
Type        &SA(s)   &de Vaucouleurs et al. 1991 \\
Diameters& 2.8$\times$1.3 arcminutes&\\
radial Velocity &1306$\pm$6 km/sec &  de Vaucouleurs et al. 1991\\
M$_B$ &  -19.82 & Tully 1988\\
L$\alpha$  &    1.08$\times$10$^{41}$erg/sec         &D=24 Mpc, Bell \& Kennicutt 2001\\
L$\alpha$ aperture size& 3 arc minutes&\\ 
\tableline
\end{tabular}
\end{table}
\clearpage 
\begin{table}
\caption{Log of the Observations \label{obser}}
\vspace*{0.4cm}
\begin{tabular}{|llll|}
\tableline\tableline
Filter &V&I&H$_{\alpha}$\\
\tableline
Exposure Time&100, 500, 500, 600&500, 500, 500&1000, 1000, 1000, 1500\\
in second& & &\\
\tableline
Date & February 5-6, 1995&&\\
\tableline
Observer& P. Notni&&\\
\tableline
\end{tabular}
\end{table}
 \clearpage
\begin{deluxetable}{lcc}
\tablewidth{0pt}
\tablecaption{Un-identified Hodge and Kennicutt (1983) HII regions \label{Hodge}}
\tablehead{
\colhead{KH-ID}           & \colhead{X-off}      &
\colhead{Y-off}}         
\startdata
  10 &   031&    010\\
  12 &   027&    009\\
  25 &  -001 &   007 \\
  32 &  -012 &  -006\\
  36 &  -017 &   017\\
  37 &  -017 &  -028\\
  40 &  -025 &   001\\
  43 &  -029 &   019\\
  44 &  -032 &   020\\
  47 &  -036 &   004\\
\enddata
\end{deluxetable}
 \clearpage
\begin{deluxetable}{lllccc}
\tablewidth{0pt}
\tablecaption{HII region in NGC 3389 \label{HII}}
\tablehead{
\colhead{Name}           & \colhead{R.A.(2000.0)}      &
\colhead{Dec. (2000.0)}          & \colhead{Diameter (pc)}  &
\colhead{log L (erg/sec)} & \colhead{HK-ID}}

\startdata

H01&10$^{h}$48$^{m}$28.80$^{s}$&12$^{d}$31$\arcmin$49.67$\arcsec$&277&38.24&1\\
H02&10  48  28.96&12  31 44.43&370&38.29&-\\
H03&10  48  28.58&12  31 48.38&472&38.74&2\\
H04&10  48  29.11&12  31 35.33&264&38.11&-\\
H05&10  48  28.99&12  31 34.54&293&37.90&3\\
H06&10  48  29.12&12  31 31.18&442&38.61&4\\
H07&10  48  28.08&12  31 51.04&815&38.97&5\\
H08&10  48  27.85&12  31 49.95&328&37.98&-\\
H09&10  48  28.41&12  31 36.13&440 &37.69&6\\
H10&10  48  27.98&12  31 41.84&279&38.03&7\\
H11&10  48  28.27&12  31 33.34&555&38.78&8\\
H12&10  48  27.18&12  31 57.46&1138&39.80&9\\
H13&10  48  26.82&12  32 00.72& 302&37.88&-\\
H14&10  48  26.33&12  32 04.66& 421&38.38&11\\
H15&10  48  26.98&12  31 42.42& 798&38.96&13\\
H16&10  48  26.52&12  31 52.30& 799&38.82&14\\
H17&10  48  26.20&12  31 59.63&591&38.41&15\\
H18&10  48  25.70&12  32 10.90&567&37.54&-\\
H19&10  48  25.97&12  32 03.48& 267&37.96&-\\
H20&0   48  25.48&12  32 14.06&642&38.17&16\\
H21&10  48  25.74&12  32 04.67& 404&38.46&17\\
H22&10  48  26.57&12  31 41.29& 165&37.70&-\\
H23&10  48  25.95&12  31 56.30& 416&38.46&18\\
H24&10  48  25.66&12  32 02.06& 490&38.63&19\\
H25&10  48  26.09&12  31 48.58& 331&38.18&-\\
H26&10  48  25.59&12  31 56.29& 724&38.82&20\\
H27&10  48  24.61&12  32 16.74& 468&38.61&22\\
H28&10  48  25.60&12  31 50.42& 549&38.33&21\\
H29&10  48  24.97&12  32 05.32& 331&37.70&-\\
H30&10  48  25.61&12  31 47.59& 452&38.91&-\\
H31&10  48  24.77&12  32 07.49& 423&38.27&23\\
H32&10  48  24.33&12  32 18.05& 198&37.80&-\\
H33&10  48  24.93&12  32 01.83& 704&39.01&24\\
H34&10  48  23.90&12  32 26.76& 495&38.44&-\\
H35&10  48  24.83&12  31 59.72& 504&38.49&-\\
H36&10  48  24.00&12  32 16.94& 457&38.57&26\\
H37&10  48  24.92&12  31 52.58& 239&37.82&-\\
H38&10  48  23.65&12  32 21.01& 358&37.90&-\\
H39&10  48  25.33&12  31 37.57& 463&38.66&27\\
H40&10  48  25.12&12  31 42.45& 396&38.23&28\\
H41&10  48  24.30&12  32 01.50& 880&38.91&29\\
H42&10  48  23.35&12  32 21.41& 564&38.67&31\\
H43&10  48  23.82&12  32 08.01&1087&39.04&30\\
H44&10  48  22.92&12  32 24.08& 320&38.32&33\\
H45&10  48  23.32&12  32 11.27& 653&38.92&35\\
H46&10  48  24.49&12  31 40.50& 285&38.05&34\\
H47&10  48  23.69&12  31 56.82& 333&37.86&-\\
H48&10  48  24.32&12  31 39.84& 291&38.36&38\\
H49&10  48  22.45&12  32 23.00& 487&38.74&39\\
H50&10  48  22.17&12  32 19.74& 442&38.09&-\\
H51&10  48  22.47&12  32 11.55& 399&38.10&-\\
H52&10  48  23.03&12  31 57.23& 460&38.40&41\\
H53&10  48  23.13&12  31 51.79& 698&39.28&42\\
H54&10  48  22.62&12  31 58.79& 270&37.70&-\\
H55&10  48  21.73&12  32 16.79& 507&38.82&45\\
H56&10  48  22.63&12  31 53.63& 618&38.99&46\\
H57&10  48  22.38&12  31 59.07& 340&38.21&49\\
H58&10  48  21.52&12  32 19.51& 308&38.08&48\\
H59&10  48  21.70&12  32 09.81& 596&38.34&-\\
H60&10  48  21.52&12  32 12.48& 400&38.10&50\\
H61&10  48  21.31&12  31 59.60& 369&38.08&-\\
\enddata
\end{deluxetable}

\end{document}